\documentclass[aoas,nameyear,dvips]{arximspdf}
\usepackage{graphicx}

\doi{10.1214/10-AOAS360}
\volume{4}
\issue{4}
\pubyear{2010}
\firstpage{2024}
\lastpage{2048}

\makeatletter
\renewcommand{\epsilon}{\varepsilon}

\newcommand{\subR}[1]{{#1}_{(R)}}
\newcommand{\bof}{\mathbf}
\newcommand\Perp{\protect\mathpalette{\protect\independenT}{\perp}}
\def\independenT#1#2{\mathrel{\rlap{$#1#2$}\mkern4.1mu{#1#2}}}
\makeatother

\begin{document}
\begin{frontmatter}

\title{A Bayesian graphical modeling approach to microRNA
regulatory network inference}
\runtitle{Bayesian networks for genomic data integration}

\begin{aug}
\author[A]{\fnms{Francesco C.} \snm{Stingo}\ead[label=e1]{stingo@ds.unifi.it}},
\author[B]{\fnms{Yian A.} \snm{Chen}\ead[label=e2]{Ann.Chen@moffitt.org}},
\author[C]{\fnms{Marina} \snm{Vannucci}\thanksref{T1}\corref{}\ead[label=e3]{marina@rice.edu}},
\author[D]{\fnms{Marianne} \snm{Barrier}\ead[label=e4]{barrier.marianne@epa.gov}}
\and
\author[E]{\fnms{Philip E.} \snm{Mirkes}\ead[label=e5]{pmirkes@gmail.com}}
\thankstext{T1}{Supported in part by NIH Grant R01-HG00331901 and NSF
Grant DMS-06-05001.}
\runauthor{F. C. Stingo et al.}
\affiliation{University of Florence and Rice University, Moffitt
Cancer Center, Rice University, Texas A$\&$M University and Texas A$\&
$M University}
\address[A]{F. C. Stingo \\
Department of Statistics\\
University of Florence \\
50134 Florence\\
Italy\\
\printead{e1}} 
\address[B]{Y. A. Chen\\
Department of Biostatistics\hspace*{5.5pt}\\
Moffitt Cancer Center\\
Tampa, Florida 33612\\
USA\\
\printead{e2}}
\address[C]{M. Vannucci\\
Department of Statistics\\
Rice University \\
Houston, Texas 77005\\
USA\\
\printead{e3}}
\address[D]{M. Barrier\\
US EPA, ORD, NHEERL\\
ISTD, SBB (MD-72)\\
Research Triangle Park,\\
\quad North Carolina 27711\\
USA\\
\printead{e4}}
\address[E]{P. E. Mirkes\\
Department of Veterinary,\\
\quad Physiology $\&$ Pharmacology \\
Texas A$\&$M University \\
College Station, Texas 77845\\
USA\\
\printead{e5}}
\end{aug}

\received{\smonth{4} \syear{2010}}
\revised{\smonth{5} \syear{2010}}

%
\begin{abstract}
It has been estimated that about 30\% of the genes in the human
genome are regulated by microRNAs (miRNAs). These are short RNA sequences
that can down-regulate the levels of mRNAs or proteins in animals
and plants. Genes regulated by miRNAs are called targets. Typically,
methods for target prediction are based solely on sequence data and
on the structure information. In this paper we propose a Bayesian
graphical modeling approach that infers the miRNA regulatory network
by integrating expression levels of miRNAs with their potential mRNA
targets and, via the prior probability model, with their
sequence/structure information. We use a directed graphical model
with a particular structure adapted to our data based on biological
considerations. We then achieve network inference using stochastic
search methods for variable selection that allow us to explore the
huge model space via MCMC. A time-dependent coefficients model is
also implemented. We consider
experimental data from a study on a very well-known developmental
toxicant causing neural tube defects, hyperthermia. Some of the
pairs of target gene and miRNA we identify seem very plausible and
warrant future investigation. Our proposed method is general and can
be easily applied to other types of network inference by integrating
multiple data sources.
\end{abstract}

%

\end{frontmatter}

\section{Introduction} One of the major tasks and challenges in the
post-genomics era is to decipher how genes and their products
(proteins) are regulated. Regulation can happen at transcriptional,
post-transcriptional, translational and post-translational level.
Transcription is the process of synthesizing a stretch of
ribonucleic acids (RNA) based on a specific DNA sequence.
Transcriptional regulation can affect whether or not a specific RNA
is transcribed as well as the amount of RNA produced. RNA can be
regulated post-transcriptionally through degradation or modification
of the RNA strand, which can affect its function. A segment of RNA
can interact with other genes or proteins or can encode a protein.
Translation, the process of forming a protein based on an RNA
sequence, can also be positively or negatively regulated. Proteins
often undergo post-translational modifications, which can affect
their function. An abundant class of short ($\sim$22 nucleotide)
RNAs, known as microRNAs (miRNAs), plays crucial regulatory roles in
animals and plants [\citet{farh2005}]. It has been estimated that at
least 30\% of the genes in human genomes are regulated by miRNAs; see
\citet{lewis2005} and also \citet{rajewsky2006}. Genes
regulated by miRNAs are generally
called ``targets.'' The actual mechanism of miRNA regulation is still
an active area of research and the complete picture of the
regulatory mechanism is still to be understood
[\citet{thermann2007}]. According to current knowledge, it is
believed that miRNAs
regulate their targets either by degrading mRNA
post-transcriptionally [\citet{bagga2005}], or by suppressing
initiation of protein synthesis [\citet{pillai2005}], and/or by
inhibiting translation elongation after initiation of protein
synthesis [\citet{petersen2006}].
The biosynthesis and maturation of
miRNAs is composed of distinctive events. Briefly, the miRNA
precursor is processed by Dicer to produce the mature,
single-stranded molecule that is incorporated into the RNA-induced
silencing complex (RISC) and possesses a 6--8 bases of ``seed''
sequence that mainly targets complementary sequences within the
3'-untranslated regions (UTR) of mRNA transcripts.

Many algorithms have been developed to predict potential target
sequences for miRNAs based on their specific sequence and structure
characteristics. These algorithms mainly use sequence information,
hybridization energy for structure prediction and cross-species
comparisons [\citet{rajewsky2006}]. Target prediction algorithms
generally take into account different factors that influence
miRNA/target interactions, such as seed match complementarity, 3'-UTR
seed match context, the conservation, favorability of free energy
binding and binding site accessibility. Some of the more widely
used algorithms include the following: TargetScan of \citet
{lewis2005}, PicTar
of \citet{krek2005}, miRanda of \citet{miranda2004}, PITA of
\citet{pita} and DIANA-microT of \citet{kiriakidou2004}. A
comprehensive review of these and other methods can be found in
\citet{yoon2006}. Typically, a large amount (e.g., hundreds to
thousands) of potential targets are predicted by these algorithms,
and it can be overwhelming for researchers to search through the
candidate targets for those which play critical regulatory roles
under particular experimental or clinical conditions.

Our goal is to develop a statistical approach to identify a small
set of potential targets with high confidence, making future
experimental validation feasible. Since miRNAs down-regulate the
expression of their targets, expression profile of miRNAs and their
potential targets can be used to infer their regulatory
relationships. We propose a Bayesian graphical modeling approach
that infers the miRNA regulatory network by integrating these two
types of expression levels. We use a directed graphical model with a
particular structure adapted to our data based on biological
considerations. We take into account current knowledge on the down-regulation
effect of mRNA expression (by miRNA) by imposing constraints on the
sign of the
regression coefficients of our proposed model.
The model also integrates the sequence/structure
information, as generated by widely used target prediction
algorithms, via the prior probability model. We then achieve network
inference using stochastic search methods for variable selection.

We consider experimental data from a study on a very well-known
developmental toxicant causing neural tube defects, hyperthermia. We
have available 23 mouse miRNAs and a total of 1297 potential
targets. We infer their regulatory network under two different
treatment conditions and also investigate time-dependent regulatory
associations. Some of the pairs of target gene and miRNA we identify
seem very plausible and warrant future investigation.

\citet{Huang2007}, \citet{Huang2008} have proposed a Bayesian model for the
regulatory process of targets and miRNAs which is similar to the one
we propose here. However, in their model formulation the authors
consider regression coefficients that are constant with respect to
the mRNAs, while our formulation allows a more efficient way of
selecting gene-miRNA pairs. Also, in order to achieve posterior
inference, we implement a full MCMC procedure while
\citet{Huang2007} adopt a variational method that only approximates
the posterior distribution. More importantly, \citet{Huang2007}
restrict their search algorithm to a preselected subset of possible
gene-miRNA relations, which they select based on the available
sequence information, therefore excluding a priori a large
number of associations that could instead occur in specific
experimental conditions, such as hyperthermia.

This paper is organized as follows. Section \ref{sec:Data}
introduces the experimental study and describes the available data,
that is, the expression data of miRNAs and their potential mRNA
targets, and the corresponding association scores. Section \ref
{sec:Model} illustrates the proposed modeling approach via a
Bayesian graphical model and describes the prior model and the
variable selection scheme. Section \ref{sec:postinf} describes how
to perform posterior inference and Section \ref{sec:appl} provides a
detailed analysis of the miRNA regulatory network reconstruction
based on the available data. Section \ref{sec:final} concludes the
paper.

\section{Neural tube defects} \label{sec:Data} Neural tube defects
(NTDs) are
some of the most common congenital defects, with approximately 12 per
day in the United States [\citet{finnell2000}]. NTDs are generally
related to failure of embryonic neural folds to fuse properly along
the neuroaxis during development. Studies in both humans and
animals suggest a complex genetic component to NTDs, likely
involving multiple loci, together with environmental factors.
MicroRNAs are believed to play important regulatory roles in mouse
development and human disease [see, for example, \citet{conrad2006}],
although detailed regulatory mechanisms are still unknown.

In this paper we consider experimental data from a study on a very
well-known developmental toxicant causing neural tube defects,
hyperthermia. In the study mice are used as the animal model to
study NTDs. Time-mated female C57Bl/6 mice were exposed in
vivo to a 10 minute hyperthermia or control treatment on
gestational day 8.5, when the neural folds are fusing to form the
neural tube. Four litters were collected for each treatment at 5,
10 and 24 hours after exposure. Each litter was treated as a
single biological sample. MiRNAs and mRNAs were extracted from each
sample for expression analysis.

\subsection{miRNA expression levels}

As the regulatory network can be very complex, we focus on a small
sets of mRNA targets with high confidence. With a limited budget
available, a pilot study was performed to screen the expression
profiles of most of the known ($\sim$240 ) mouse microRNAs based on
one set of samples, for both heat shock and control at 4 different
time points, and using TaqMan miRNA RTPCR assays available at the
time (Applied Biosystems, Foster City, CA; provided in collaboration
with Ambion, Austin, TX).
Of the 240 miRNA evaluated, 50 had none or very low expression at
all time points, while 86 had a 2-fold or greater change in
expression in response to hyperthermia exposure at one or more time
point. From this set of 86 miRNA, we chose a subset of 23 miRNA
whose patterns of expression were interesting enough for further
analysis and obtained replicate sample sets. The complete experiment
was therefore carried out using only this set of 23 miRNAs.
Results from the analysis of these data, and their biological interpretation,
are clearly limited by this initial choice.

MicroRNA was extracted from each sample at each time point under
each experimental condition. Two technical replicates were prepared
for RTPCR quantification to confirm the technical reproducibility.
In RTPCR experiments, fluorescence techniques are used to detect the
amplification of miRNAs to assess their abundance. A fluorescence
threshold is determined for an experiment, and the cycle number,
which reaches the predetermined threshold level of log2-based
fluorescence, is defined as the Ct number.
An inverse linear relationship exists between Ct number and the
logarithm of input quantity of the gene when the amplification
efficiency is perfect [\citet{pfaffl2001}]. The Ct numbers of the
miRNA technical replicates were averaged across the two technical
replicates.

\subsection{Target prediction via sequence and structure information}
Four of the most widely used algorithms, miRanda, TargetScan, PITA
and PicTar, were used in our study to retrieve the sequence and
structure information for target prediction. The algorithm miRanda
[see \citet{miranda2004} and \citet{enright}] computes optimal
sequence complementarity between mature microRNAs and a mRNA using a
weighted dynamic programming algorithm with weights that are
position-dependent and that reflect the relative importance of the
$5^\prime$ and $3^\prime$ regions. Its alignment score is a weighted sum of match
and mismatch scores for base pairs and gap penalties. The
free-energy of the formation of the microRNA:mRNA duplex is used by
miRanda as a filtering step. PITA of \citet{pita} focuses on the
overall effect of all potential binding sites combined together on
the given UTR. Pictar [see \citet{krek2005}] utilizes genome-wide
alignment among species to take conservation into consideration.
Finally, TargetScan of \citet{lewis2005} utilizes two orthogonal
scores, one is the total context score, and the other independent
score is the probability of conserved targeting. More details on
each algorithm can be found in the original papers.

The prediction scores using the four algorithms, miRanda, TargetScan, PITA
and PicTar, can be obtained from the respective websites. Predictions
by PicTar were obtained
from the PicTar
site\setcounter{footnote}{1}\footnote{\url{http://pictar.mdc-berlin.de/}.}. A zero or absent PicTar score indicates that the raw
score did not exceed a prespecified threshold, that is, the
algorithm suggests no indication of a regulatory association. The
current release (September 2009) comprising 1,209,841 predicted microRNA
target sites in 26,697 mouse gene isoforms for 491 mouse miRNAs,
generated by the miRanda algorithm of \citet{miranda2004}, was
downloaded from \texttt{MICRORNA.ORG}; see \citet{betel2008}. PITA
Catalog version 6 (31-Aug-08) was downloaded from Segal lab's
website\footnote{\url{http://genie.weizmann.ac.il/pubs/mir07/mir07_data.html}.}.
The target scores from all predictions were used in our study.
TargetscanMouse 5.1 (released April 2009) was downloaded
from the
internet\footnote{\url{http://www.targetscan.org/cgi-bin/targetscan/data_download.cgi?db=mmu\_50}.}. Scores for preferential conservation of
the sites (Aggregate PCT) and the context of the sites within the
UTR (total context score) were parsed and used in our study. Matlab
scripts were written to retrieve the prediction scores from each
algorithm using the RefSeq Ids of all potential targets downloaded
for the 23 mouse miRNAs of interest.

\subsection{Target mRNA expression levels}

RNA was extracted from each sample at each time point and hybridized
to GE Codelink Mouse Whole Genome Microarrays (GE Healthcare Life
Sciences, Piscataway, NJ). The slides were scanned and mRNA
expression levels were quantified. One biological sample was not
prepared properly at hour 10 in the control group, and therefore
discarded.

The RefSeq Ids of the probes spotted on the Codelink microarrays
were linked to the retrieved potential targets of the 23 miRNAs
previously identified. The mRNAs were included in the analysis only
if they were among the potential targets predicted by the four
algorithms described above. Genes with missing or negative values were
excluded from the analysis. The expression levels of the remaining
mRNAs were then log2 transformed so that both miRNA and mRNA
expression were on the log2 scale. A total of 1297 potential
targets were included in the final analysis. The transformed
expressions across the 3 time points were centered by subtracting
their means.

\section{Model}\label{sec:Model} We have available expression levels on a set of
miRNAs and their potential targets. For each target we are
interested in identifying a small number of regulatory associations
with high confidence. We have also available sequence information
for target prediction in the form of scores of regulatory
associations. We propose a Bayesian graphical modeling approach that
infers the miRNA regulatory network by integrating the expression
data and, via the prior probability model, the sequence/structure
information. An important aspect of our methodology is the concept
of sparsity, that is, we believe that most genes are regulated by a
small number of miRNAs.

\subsection{A Bayesian network for gene \& miRNA expression}

\begin{figure}

\includegraphics{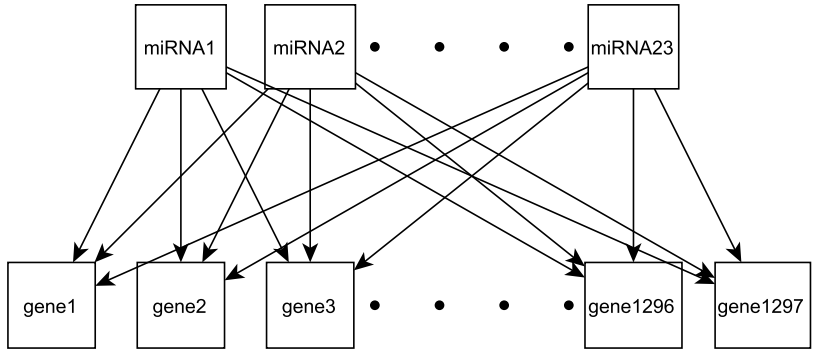}

\caption{Graphical representation of the miRNA regulatory network.}\label{Rnet}
\end{figure}

We use a directed graphical model (Bayesian Network) with a
particular structure adapted to our data that uses a predetermined
ordering of the nodes based on biological considerations. This model
is able to answer to the baseline question of ``\textit{which miRNAs
regulate which targets}'' and, in addition, allows us to build a fast
computational procedure required in such a high-dimensional
framework. A graphical representation of the full miRNA network is
given in Figure \ref{Rnet}. Our task is to find a significant subset
of edges.

Graphical models are graphs in which nodes represent random
variables and the lack of arcs represents conditional independence
assumptions; see, for example, \citet{cowell1999}. Graphical models
provide a compact representation of joint probability distributions.
Here we work with a multivariate normal distribution, and therefore
with a Graphical Gaussian model (GGM). A graph $\mathcal{G}$ and the
covariance matrix $\Omega$ entirely define a GGM $\mathcal{M}$,
$\mathcal{M} \equiv(\mathcal{G}, \Omega)$. Arcs can be undirected,
indicating symmetric dependencies, or directed, when there is a
direction of the dependence. These dependencies can come from prior
knowledge or from data analysis. Undirected graphical models have a
simple definition of independence, for example, two nodes A and B are
conditionally independent given a third set, C, if all paths between
the nodes in A and B are separated by a node in C. Directed
graphical models need a specific ordering of the variables. Graphs
that do not allow the presence of cycles are called directed acyclic
graphs (DAG). Conditional independencies in a DAG depend on the
ordering of the variables.

We work with a DAG and impose an ordering of the variables such that
each target can be affected only by the miRNAs and that the miRNAs
can affect only the targets. Let ${\bof Z} = ({\bof Y}_1,{\bof
Y}_2,\ldots,{\bof Y}_G,{\bof X}_1,\ldots,{\bof X}_M)$ with ${\bof
Y}=({\bof Y}_1,\ldots,{\bof Y}_G)$ the matrix representing the targets
and ${\bof X}=({\bof X}_1,\ldots,{\bof X}_M)$ the miRNAs .
Specifically, $y_{ng}$ indicates the normalized averaged $\log_2$
gene expression of gene $g = 1,\ldots,G$ in sample $n = 1, \ldots,
N$. These expression values are biological replicates obtained by
averaging two technical replicates. Similarly, $x_{nm}$ indicates
the expression of the $m$th miRNA in sample $n$, with $m =
1,\ldots,M$. We have $G=1297$ and $M=23$. In addition, we have
$N=11$ i.i.d. observations under the control status and $N=12$
i.i.d. observations under hyperthermia. We infer the miRNA
regulatory network separately under the two conditions.

Our assumptions are that ${\bof Z}$ is a matrix-variate normal
variable with zero mean and a variance matrix $\Omega$ for its
generic row, that is, following \citet{Dawid1981} notation,
\[
{\bof Z} -{\bof0} \sim\mathcal{N}( I_N , \Omega).
\]
In addition, we assume that the target genes are independent
conditionally upon the miRNAs, that is, ${\bof Y}_i
\Perp{\bof Y}_j | {\bof X}_1,\ldots, {\bof X}_M$ and,
without loss of generality, that the miRNAs are independent, that
is, ${\bof X}_i \Perp{\bof X}_j$. Note that the
marginal distribution of $({\bof X}_1,\ldots,{\bof X}_M)$ does not
affect the regulatory network. In a Bayesian Network framework these
assumptions imply an ordering of the nodes and, consequently, a
likelihood factorization of the type:
%
\begin{equation}\label{model}
 f({\bof Z}) = \prod_{g=1}^G f({\bof Y}_g|{\bof X})
\prod_{m=1}^M f({\bof X}_m),
\end{equation}
where $f({\bof Y}_g|{\bof X}) \sim N({\bof X} \beta_g, \sigma_g I_N)$
and $f({\bof X}_m) \sim N(0,\sigma_m I_N)$, with $\beta_g=\break
\Omega_{\bof {XX}}^{-1} \Omega_{{\bof X}{\bof Y}_g}$ and $\sigma_g =
\omega_{gg} - \Omega_{{\bof X}{\bof Y}_g}^T \Omega_{\bof {XX}}^{-1}
\Omega_{{\bof X}{\bof Y}_g}$. Here $\omega_{gg}$ indicates the $g$th
diagonal element of $\Omega$ and $\Omega_{\bof {XX}}$, $\Omega_{{\bof
X}{\bof Y}}$ are the blocks of the covariance matrix according to the
following partition:
\[
\Omega=
\pmatrix{
\Omega_{\bof {YY}}&\Omega_{\bof {YX}}\cr
\Omega_{\bof {XY}}&\Omega_{\bof {XX}}
}.
\]
For $m=1,\ldots,M$ we have $\sigma_m=\omega_{mm}$.

\begin{figure}

\includegraphics{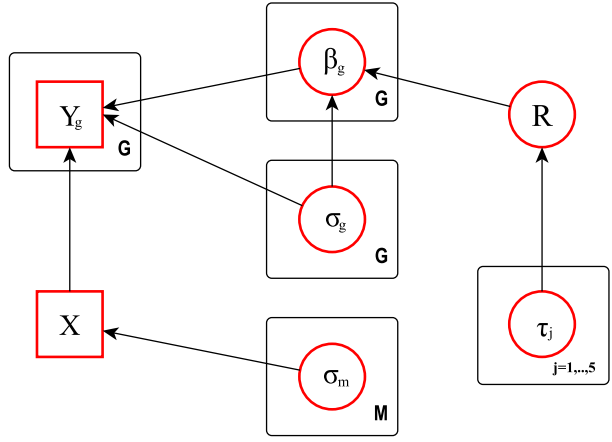}

\caption{Structure of the graphical model.}\label{fig1}
\end{figure}

According to current knowledge, miRNAs down-regulate gene
expression. It therefore seems appropriate to include this
information into our statistical model. This is achieved by
specifying negative regression coefficients
via the prior model.
First, we note that our model is equivalent to the following system
of equations:
%
\begin{equation}\label{eqns}
\cases{
{\bof Y}_{1}= -{\bof X}\beta_{1} + \epsilon_{\sigma_1}, \cr
\hspace{4.5pt}\vdots \cr
{\bof Y}_{G}= -{\bof X} \beta_{G}+\epsilon_{\sigma_G},
}
\end{equation}
where $\epsilon_{\sigma_g}$ is distributed as a multivariate normal
with zero mean and covariance matrix $\sigma_g I_N$. Then, we
complete the model specification by specifying prior distributions
on the regressions coefficients and the error variances. We impose our
biological constraints by using Gamma distribution priors for the
positive regressions coefficients, $(\beta_{gm}|\sigma_g) \sim \mathit{Ga}(1,
c  \sigma_g)$, and Inverse-Gamma distributions for the error
variances, $\sigma_g^{-1} \sim \mathit{Ga}((\delta+M)/2, d/2)$. Figure
\ref{fig1} shows a graphical representation of our model. Circles
indicate parameters and squares observed random variables. The
parameters $\bof{R}$ and $\tau$ are involved in the variable
selection and are introduced in the section below.

\subsection{Prior model for variable selection}

The goal of the analysis is to find, for each target, a small subset
of miRNAs that regulate that target with high probability. This can
be framed into a variable selection problem. Specifically, we can
introduce a $(G\times M)$ matrix $\bof{R}$ with elements $r_{gm}=1$
if the $m$th miRNA is included in the regression of the
$g$th target and $r_{gm}=0$ otherwise. Conditioned upon
$\bof{R}$, expression (\ref{eqns}) is equivalent to a system of linear
equations where the included regressors are only those miRNAs
corresponding to $r_{gm}=1$. To emphasize the variable selection
nature of our model, we write it as follows:
\begin{equation}\label{modelR}
\mathbf{Y}_g = -\subR{\mathbf{X}} \subR{\beta_g} + \epsilon_{\sigma_g},
\end{equation}
where $\subR{\beta_g}$ is the vector that is formed by taking only
the nonzero elements of $\beta_g$ and $\subR{\mathbf{X}}$ is the
matrix that is formed by taking only the corresponding columns of~$\mathbf{X}$.
The goal of our modeling is to infer which elements of the vectors
$\beta_g$'s are non-zero, indicating a relationship between the
corresponding genes and miRNAs. This underlying regulatory network
is encoded by the association matrix $\bof{R} = \{ r_{gm}\}$.
 The elements of the vectors $\beta_g$'s are then
stochastically independent, given the regulatory network $\bof{R}$,
and have the following mixture prior distribution:
%
\begin{equation}\label{prior}
\pi(\beta_{gm}|\sigma_g, r_{gm}) = r_{gm} \mathit{Ga}(1,
c  \sigma_g) + (1-r_{gm}) I_{[\beta_{gm}=0]}.
\end{equation}
In addition, taking into account the regulatory network, we obtain
that $\sigma_g^{-1}| \bof{R} \sim Ga((\delta+k_g)/2, d/2)$, where
$k_g$ is the number of significative miRNAs in the regression of the
$g$th target.

Mixture priors have been used extensively for variable selection in
linear regression settings; see \citet{george1993} for univariate
regression and \citet{brown1998} and \citet{sha2004} for
multivariate models. According to prior (\ref{prior}), when
$r_{gm}=0$, then $\beta_{gm}$ is estimated by $0$ and the
corresponding column of $\mathbf{X}$ is excluded from the
$g$th equation in model (\ref{eqns}). Notice that the
dimensions of the matrix $\mathbf{X}$ are such that there are many
more columns than rows. In the domain of classical regression, this
results in insufficient degrees of freedom to fit the model unless
constraints are placed on the regression coefficients $\beta_g$'s.
Conversely, this problem is readily addressed in the Bayesian
paradigm and is known as the ``small $n$, large $p$'' framework. The
variable selection formulation we adopt here overcomes the somehow
rigid structure of the model in \citet{brown1998}, which does not
allow to select different predictors for different responses. See
also \citet{mahlet2009} for an approach based on partition models.

\subsection{Using association scores in the prior model}\label{subsec:pictarmodel}

Scores of possible associations between gene-miRNA pairs obtained
from sequence/structure information were used to estimate prior
probabilities of miRNAs binding to their target genes. Let $s_{gm}$
denote a generic score for gene $g$ and miRNA $m$, obtained, for
example, by the PicTar algorithm. As previously described, $s_{gm}$
is either positive or, in the case of a regulatory association that
is believed to be absent, equal to zero. Also, the PicTar algorithm
shrinks small values to zero, setting $s_{gm}=0$ if
$s_{gm}<\xi$ where $\xi$ is a prespecified
threshold used by the algorithm. In our model the Bernoulli random
variable $r_{gm}$ indicates whether there is a relationship between
gene $g$ and miRNA $m$. We choose to model the success probability
of $r_{gm}$ as a function of the $s_{gm}$ score as follows:
\begin{equation}\label{bernprior}
 P(r_{gm} = 1| \tau) = \frac{\exp[\eta+\tau
s_{gm}]}{1 + \exp[\eta+\tau s_{gm}]},
\end{equation}
where $\tau$ is an unknown parameter. We then assume that the
elements of $\bof{R}$ are stochastically independent given $\tau$.
Notice that for $s_{gm} = 0$, we have that $P(r_{gm}=1) =
\exp[\eta]/(1+\exp[\eta])$, which gives a 0.5 prior probability
when $\eta=0$. Thus, the inverse logit transformation
of $\eta$ can be interpreted as the false negative rate associated
with the PicTar thresholding scheme. For a score $s_{gm} > 0$ we
have $P(r_{gm}=1) > \eta$, with higher scores yielding higher prior
probabilities of association. We further specify a hyperprior on
$\tau$ as a gamma distribution $\tau\sim \mathit{Ga}(a_{\tau}, b_{\tau})$,
ensuring the positivity of the parameter.

Since we have available multiple prior sources of information, from
different sequence/structure algorithms, it makes sense to combine
them all by incorporating all scores into the prior distribution
via additional $\tau$ parameters. For example, in the application
of Section \ref{sec:appl} we combine five different scores as
\[
P(r_{gm} = 1| \tau) =
\frac{\exp[\eta+\tau_1 s^1_{gm} + \tau_2 s^2_{gm}+\tau_3 s^3_{gm}
+ \tau_4 s^4_{gm}+\tau_5 s^5_{gm}]}{1
+ \exp[\eta+\tau_1 s^1_{gm} + \tau_2 s^2_{gm}+\tau_3 s^3_{gm}+
\tau_4 s^4_{gm}+\tau_5 s^5_{gm}]},\label{bernpriorM}
\]
with $\tau=(\tau_1,\ldots,\tau_5)$ and where the ${s^j_{gm}}$'s,
with $j=1,\ldots,5$, denote the PicTar, miRanda, aggregate Target
Scan, total Target Scan and PITA scores, respectively. Scores should
be normalized to obtain positive values that lie in the same range,
with bigger values corresponding to stronger prior connections.

\subsection{Time-dependent coefficients model}

The previous model implies that the relation between gene $g$ and
miRNA $m$ is constant over time. In the experimental study for
which we developed our model there is no dependence between the
measurements at different time points, since these observations
come from independent units. However, one may still wish to
incorporate into the model the fact that relations may possibly
change with time. This can be done by allowing different regression
coefficients at different time points, as follows:
\begin{equation}\label{modelTime}
\cases{
 \mathbf{Y}_{1}=-\mathbf{X}\beta_{1}-\mathbf{X}_2^*\beta_{1}' - \mathbf{X}_3^* \beta_{1}'' +
 \epsilon_{\sigma_1},\cr
 \hspace{4.5pt}\vdots\cr
 \mathbf{Y}_{G}= -\mathbf{X}\beta_{G} - \mathbf{X}_2^* \beta_{G}' -\mathbf{X}_3^* \beta_{G}'' + \epsilon_{\sigma_G}, \cr
}
\end{equation}
where the $\mathbf{Y}_g$'s are $N \times1$ vectors and
\begin{eqnarray*}
 \mathbf{X} =
 \pmatrix{
    \mathbf{X}_1 \cr
    \mathbf{X}_2 \cr
    \mathbf{X}_3
},
\qquad
\mathbf{X}_2^* =
\pmatrix{
 \mathbf{0} \cr
 \mathbf{X}_2 \cr
\mathbf{0} \cr
},
\qquad
 \mathbf{X}_3^* = \pmatrix{
\mathbf{0} \cr
 \mathbf{0} \cr
 \mathbf{X}_3
},
\end{eqnarray*}
are the $N \times M$ matrices of the observed values, with
$\mathbf{X}_1$, $\mathbf{X}_2$ and $\mathbf{X}_3$ the miRNA
expressions collected at the first, the second and the third time
point, respectively. The element $\beta_{gm} \in\beta_g$
represents the relation between gene $g$ and miRNA $m$ at the first
time point, $\beta_{gm}+\beta_{gm}'$, with $\beta_{gm}' \in
\beta_g'$, represents the relation at the second time point and
$\beta_{gm}+\beta_{gm}''$, with $\beta_{gm}'' \in\beta_g''$, at the
third time point.

In order to do variable selection on the elements of $\beta_g'$ and
$\beta_g''$, we introduce two additional binary matrices $\bof{R}'$
and $\bof{R}''$, with a similar role to $\bof{R}$ in the
time-invariant model (\ref{modelR}). We consider the elements of
$\bof{R'}$ and $\bof{R}''$ independently distributed and following a
Bernoulli distribution with parameter $P(r_{gm}'=1) = \eta_b =
P(r_{gm}''=1) $. Because of the way we implement the MCMC (see
Section \ref{sec:postinf}), we do not need to impose the sequence
information on the prior on $\bof{R}'$ and $\bof{R}''$.

As for the elements of the $\beta_g$'s vectors, we assume that the
elements of the $\beta_g'$'s and $\beta_g''$'s vectors are
stochastically independent given the regulatory networks $\bof{R}'$
and $\bof{R}''$, respectively, and that they have the following
prior distributions:
\begin{eqnarray*}
\pi(\beta_{gm}'|\sigma_g, r_{gm}')&=&r_{gm}' N(0,c^{-1}\sigma_g \zeta)
+ (1-r_{gm}') I_{[\beta_{gm}'=0]},\\
\pi(\beta_{gm}''|\sigma_g,r_{gm}'')&=&r_{gm}'' N(0,c^{-1}\sigma_g
\zeta)+(1-r_{gm}'') I_{[\beta_{gm}''=0]},
\end{eqnarray*}
where the hyperparameter $\zeta$, usually $\leq1$, reflects the
prior information on the magnitude of the $\beta_g'$'s and
$\beta_g''$'s.

We can reframe the time-dependent coefficients model in the same way
we have framed model (\ref{modelR}), that is,
\[
\mathbf{Y}_g = -\subR{\mathbf{X}} \subR{\beta_g} -
\mathbf{X}_{2(R')}^* \beta_{g(R')}' - \mathbf{X}_{3(R'')}^*
\beta_{g(R'')}'' + \epsilon_{\sigma_g},
\]
where the columns of $\mathbf{X}_2^*$ are selected if the
corresponding elements of $\bof{R}'$ are equal to 1 and the columns
of $\mathbf{X}_3^*$ are selected if the corresponding elements of
$\bof{R}''$ are equal to 1, for each equation.

\section{Posterior inference} \label{sec:postinf}

For posterior inference the primary interest is in estimating the
association matrix $\bof{R}$. Here we show that $\bof{R}$ can be
estimated by designing a simple extension of the stochastic search
procedures used for variable selection; see \citet{george1993} and
\citet{sha2004}, among many others.

We use a Metropolis--Hastings within Gibbs to explore the huge model
space and find the most influential predictors. Our model has 23
regressors for each of 1297 equations, that is a total of 29,831
regression coefficients for the time invariant model (\ref{modelR})
and 89,493 for the time dependent model (\ref{modelTime}).
Clearly, exploring such a huge posterior space is challenging. Here
we exploit the sparsity of our model, that is, the belief that most of
the genes are well predicted by a small number of regressors, and
resort to a Stochastic Search Variable Selection (SSVS) method. A
stochastic search allows us to explore the posterior space in an
effective way, quickly finding the most probable configurations,
that is, those corresponding to the coefficients that have high
marginal probability of $r_{gm}=1$, while spending less time in
regions with low posterior probability.

In order to design this MCMC search, we need to calculate the
marginal posterior distribution of $\bof{R}$ by integrating out
$\beta_g$ from the posterior:
\begin{eqnarray*}
&&f\bigl(\mathbf{Y}_g|\subR{\mathbf{X}}, \sigma_g, \bof{R}\bigr)  \propto
\frac{1}{(2 \pi)^{(N-k_g)/2} \sigma_g^{N/2}c^{k_g}} |U_g|^{1/2} \\
&&\hphantom{f\bigl(\mathbf{Y}_g|\subR{\mathbf{X}}, \sigma_g, \bof{R}\bigr)\propto}
\displaystyle{}\times\exp\biggl[\frac{1}{2 \sigma_g} q_g \biggr] \Phi_{k_g}(0;-U_g C_g, \sigma_g
U_g),
\end{eqnarray*}
where $U_g = (\subR{\mathbf{X^T}}\subR{\mathbf{X}})^{-1}$, $C_g =
\mathbf{Y}_g^T \subR{\mathbf{X}^T} - (\sigma_g^{1/2}/c)
\mathbf{1}_{k_g}$ and $q_g = \mathbf{Y}_g^T \mathbf{Y}_g - C_g U_g
C_g^T$ and with $k_g$ the number of selected regressors. Here
$\Phi_{k_g}(0;-U_g C_g,\break\sigma_g U_g)$ indicates the cdf of a
multivariate normal, with mean $-U_g C_g$ and covariance matrix
$\sigma_g U_g$, calculated at the zero vector.

Our algorithm consists of three steps. The first step is based on
the marginal posterior distribution conditioned upon $\tau_1, \ldots,
\tau_5$ and $\sigma_g$ and consists of either the addition or the
deletion of one arrow in our graphical model or the swapping of two
arrows. The second step generates new values of $\tau_j$'s from their
posterior distribution. In the last step values
of all the error variances $\sigma_g$ are updated. The un-normalized
full conditionals needed for the Gibbs sampler can be derived from
the conditional independencies of our model, as given in Figure
\ref{fig1}. We now describe the three steps of the algorithm:

\begin{enumerate}
\item We use one of two types of moves to update $\bof{R}$:
\begin{itemize}
\item with probability $\phi$, we add or delete an element
by choosing at random one component in the current $\bof{R}$
and changing its value;
\item with probability $1-\phi$, we swap two elements
by choosing independently at random one 0 and one 1 in the
current $\bof{R}$ and changing the value of both of them.
\end{itemize}

The proposed $\bof{R}^{\mathrm{new}}$ is then accepted with a probability that
is the
ratio of the relative posterior probabilities of the new versus the current
model:
%
\begin{equation} \label{MC1}
\min\biggl[ \frac{f(\mathbf{Y}|\mathbf{X}_{(R^{\mathrm{new}})}, \bof
{R}^{\mathrm{new}}, \sigma_g)
\pi(\bof{R}^{\mathrm{new}}|\tau)}{f(\mathbf{Y}|\mathbf{X}_{(R^{\mathrm{old}})},
\bof{R}^{\mathrm{old}}, \sigma_g)
\pi(\bof{R}^{\mathrm{old}}| \tau)},1\biggr].
\end{equation}
Because these moves are symmetric, the proposal distribution does
not appear in the previous ratio.

\item In order to update the $\tau_j$'s, we employ Metropolis steps.
The proposal is made via a truncated normal random walk kernel. The
proposed $\tau_j^{\mathrm{new}}$ is
then accepted with probability
\begin{equation}
\min\biggl[ \frac{\pi(\bof{R}| \tau_j^{\mathrm{new}}) \pi(\tau_j^{\mathrm{new}})
q(\tau_j^{\mathrm{old}};\tau_j^{\mathrm{new}})}{\pi(\bof{R}| \tau_j^{\mathrm{old}}) \pi(\tau_j^{\mathrm{old}})
q(\tau_j^{\mathrm{new}};\tau_j^{\mathrm{old}})},1 \biggr],
\end{equation}
where $q(\tau_j^{\mathrm{old}};\tau_j^{\mathrm{new}})$ is a truncated normal
with mean
$\tau_j^{\mathrm{new}}$ and truncation at 0, given the constraint of positivity on
$\tau_j$. The variance of this distribution represents the tuning
parameter and
has to be set in such a way to explore the parameter space and have a good
acceptance rate; see also Section \ref{sec:appl}.

\item For $g=1,\ldots,G$, we update the error variance $\sigma_g$
using a Metropolis step where the proposal distribution
$q(\sigma_g^{\mathrm{old}};\sigma_g^{\mathrm{new}})$ is a Gamma distribution with
parameters $a_{\sigma}$ and $b_{\sigma}$. The proposed new value is
then accepted with probability
\begin{equation}
\min\biggl[ \frac{f(\mathbf{Y}|\mathbf{X}_{(R)}, \bof{R}, \sigma
_g^{\mathrm{new}}) \pi(\sigma_g^{\mathrm{new}})
q(\sigma_g^{\mathrm{old}};\sigma_g^{\mathrm{new}})}{f(\mathbf{Y}|\mathbf{X}_{(R)},
\bof{R}, \sigma_g^{\mathrm{old}}) \pi(\sigma_g^{\mathrm{old}})
q(\sigma_g^{\mathrm{new}};\sigma_g^{\mathrm{old}})},1 \biggr].
\end{equation}
To obtain an efficient exploration of the parameter
space we set $a_{\sigma} = \sigma_g^{\mathrm{old}}/b_{\sigma}$ and
$b_{\sigma} = e / \sigma_g^{\mathrm{old}}$, where $e$ represents the
variance of the proposal distribution and can be set to obtain
a suitable acceptance ratio.

\end{enumerate}

Posterior inference can then be performed based on the MCMC output
using the marginal probabilities of the singles $r_{gm}$'s.

The MCMC algorithm for the time-dependent coefficient model
(\ref{modelTime}) is pretty similar to the procedure described
above, the main difference being that at the first step we update
either $\bof{R}$, $\bof{R}'$ or $\bof{R}''$. We then derive the
marginal posterior distribution $f(\mathbf{Y}_g|\subR{\mathbf{X}},
\bof{R})$ for the time dependent model obtaining
\begin{eqnarray*}
&&\displaystyle f\bigl(\mathbf{Y}_g|\subR{\mathbf{X}},\mathbf{X}_{2(R')}^*,\mathbf{X}_{3(R'')}^*,
\bof{R}, \bof{R}', \bof{R}'', \sigma_g\bigr) \\
 &&\qquad\displaystyle =(2\pi)^{-(n-k_g)/2} \sigma_g^{-n/2}
c^{-k_g-(k_{2g}+k_{3g})/2}\zeta^{-(k_{2g}+k_{3g})/2} |A_g|^{-1/2}|C_g|^{-1/2} \\
&&\qquad\quad\displaystyle{}\times|E_g|^{-1/2} \exp\biggl[\frac{1}{2 \sigma_g}q_g \biggr]\Phi_{k_g}(0;-E_g^{-1} F_g, \sigma_g E_g^{-1}),
\end{eqnarray*}
with
\begin{eqnarray*}
\hspace*{-30pt}q_g & = & \mathbf{Y}_g^T \mathbf{Y}_g - \mathbf{Y}_{2g}^T
\mathbf{X}_{2 (R')} A_g^{-1} \mathbf{X}_{2(R')}^{T} \mathbf{Y}_{2g}
- \mathbf{Y}_{3g}^T \mathbf{X}_{3 (R'')} C_g^{-1}
\mathbf{X}_{3(R'')}^{T} \mathbf{Y}_{3g}\\
\hspace*{-30pt}&&{}- F_g^T E_g^{-1} F_g,
\end{eqnarray*}
\begin{eqnarray*}
F_g & = & -\mathbf{X}^{T}_{(R)} \mathbf{Y}_g + \mathbf{X}_{3
(R)}^{T} \mathbf{X}_{3(R'')} C_g^{-1} \mathbf{X}_{3(R'')}^{T}
\mathbf{Y}_{3g} + \mathbf{X}_{2(R)}^{T} \mathbf{X}_{2 (R')} A_g^{-1}
\mathbf{X}_{2(R')}^{T} \mathbf{Y}_{2g} \\
& & - \sigma_g^{1/2} c^{-1} \mathbf{1}_{k_{g}}, \\
E_g & = & \mathbf{X} \mathbf{X}^{T} - \mathbf{X}_{2 (R)}^{T}
\mathbf{X}_{2(R')} A_g^{-1} \mathbf{X}_{2 (R')}^{T} \mathbf{X}_{2
(R)} - \mathbf{X}_{3 (R)}^{T} \mathbf{X}_{3(R'')} C_g^{-1}
\mathbf{X}_{3 (R'')}^{T} \mathbf{X}_{3 (R)},\\
A_g &=& \bigl(\mathbf{X}_{2 (R')}^{T} \mathbf{X}_{2(R')} + (c \zeta)^{-1}
I_{k_{2g}}\bigr), \\
C_g &=& \bigl(\mathbf{X}_{3 (R'')}^{T} \mathbf{X}_{3(R'')} + (c \zeta
)^{-1} I_{k_{3g}}\bigr)
\end{eqnarray*}
and $\mathbf{Y}_g^T = (\mathbf{Y}_{1g}^T,
\mathbf{Y}_{2g}^T, \mathbf{Y}_{3g}^T)$; $k_{2g}$ and $k_{3g}$ are
the number of selected $\beta_{gm}'$ and $\beta_{gm}''$. We can now
write the first step of the MCMC as follows:
\begin{itemize}[$1'.$]
\item[$1'.$] We first select which of the three matrices to update.
We choose $\bof{R}$ with probability $\lambda$ and
$\bof{R}'$ (or $\bof{R}''$) with probability
$(1-\lambda)/2$. We then use the same add/delete or swap
scheme described above and we accept the proposed
$\bof{R}^{\mathrm{new}}$ (or $\bof{R}'^{\mathrm{new}}$ or $\bof{R}''^{\mathrm{new}}$).
For $\bof{R}$ the acceptance probability is
\begin{eqnarray*}\label{MC2}
\min\biggl[ \frac{f(\mathbf{Y}|\mathbf{X}_{(R^{\mathrm{new}})},
\mathbf{X}_{2(R'^{\mathrm{old}})}^*,\mathbf{X}_{3(R''^{\mathrm{old}})}^*,
\bof{R}^{\mathrm{new}}, \bof{R'}^{\mathrm{old}}, \bof{R''}^{\mathrm{old}}) \pi(\bof{R}^{\mathrm{new}}|
\tau)}
{f(\mathbf{Y}|\mathbf{X}_{(R^{\mathrm{old}})},\mathbf{X}_{2(R'^{\mathrm{old}})}^*,
\mathbf{X}_{3(R''^{\mathrm{old}})}^*,
\bof{R}^{\mathrm{old}}, \bof{R'}^{\mathrm{old}}, \bof{R''}^{\mathrm{old}}) \pi(\bof{R}^{\mathrm{old}}|
\tau)},1 \biggr]
\end{eqnarray*}
and similarly if $\bof{R}'$ or $\bof{R}''$ is selected.
Note that to perform this step we need to use
only the prior distribution of the selected matrix.
\end{itemize}

This algorithm can be run either without any constraint on the moves
relative to $\bof{R}$, $\bof{R}'$ and $\bof{R}''$ or with the
constraint that the elements of $\bof{R'}$ (or $\bof{R}''$) can be
selected only when the corresponding element of $\bof{R}$ is already
selected and that the elements of $\bof{R}$ can be eliminated only
when the corresponding element of $\bof{R}'$ and $\bof{R}''$ are not
selected. For our application we adopted the constraint strategy. To
implement this, we do not need to add the ratio of the proposal
distributions into (\ref{MC1}), since we use symmetric moves. This
choice, jointly with some empirical results (not reported here), led
us to not use association scores into the prior
distribution of $\bof{R}'$ and $\bof{R}''$, because the selecting
constraints imply that the prior probability of selecting the
generic element $r_{gm}'$ (or $r_{gm}''$) already depends on the
association scores information through the prior probability on the
corresponding element $r_{gm}$. This also implies a faster
computational procedure in comparison with the option of including
the external information into the prior of $\bof{R}'$ and
$\bof{R}''$.

\section{Neural tube defects application} \label{sec:appl}

We now apply our model to analyze the data described in Section
\ref{sec:Data}, combining miRNA and mRNA expression levels with
sequence information. Our model allows us to identify significant
miRNAs for each target, possibly along the time.

\subsection{Parameter settings}

We first need to set the values of the hyperparameters of the model.
The parameter $c$ of the prior distribution of the regression
coefficients $\beta_{gm}$ can be interpreted as a correction factor.
Since truncating at zero a zero mean normal distribution with variance
$\sigma^2$ results in a half normal distribution with variance
$\approx0.7 \sigma^2$, we decided to set $c=0.7$. Also, we specified
a vague prior on $\sigma_g$ by setting $\delta=3$, the minimum value
such that the expectation of $\Omega$ exists, and chose $c = 0.2$,
setting the expected value of the variance parameter $\sigma_g$
comparable in size to a small percentage of the expected error
variances of the standardized $\mathbf{Y}$ given $\mathbf{X}$.

In our variable selection framework, the parameter $\eta$ of the
Bernoulli distribution (\ref{bernprior}) reflects the prior belief
about the percentage of significant coefficients in the model. In
this application, having 23 regressors for each of the 1297
equations, we set $\eta= -3$ to obtain a prior expected number of
regressors approximately equal to 1. This setting also corresponds to a
5\%
prior probability of selection. In Section \ref{sec:res} we show
that, even though $\eta$ affects the number of selected coefficients,
no sensitivity to the posterior selection of the most influential arcs
is observed.
For the more computationally expensive
time dependent model, we set $\eta=-3$ and $\eta_b=0.05$, to avoid
memory problems. We also set the hyperparameters $a_{\tau}=1.5$ and
$b_{\tau}=0.2$ to obtain a Gamma distribution that gives high
probability to a broad set of values of $\tau_1, \ldots, \tau_5$,
taking into account the scale of values that come from PicTar and
the other algorithms. However, the posterior distributions we obtained,
in all the different chains we ran, showed that the parameter
setting of the Gamma distribution is not strongly informative. When
running MCMCs we have set
the variance of the truncated normal proposal distribution of
$\tau_j$ equal to $0.01$ to obtain an acceptance rate
close to $25\%$.

We ran two different chains for each of the four possible
combinations, the time invariant model for the control and the
hyperthermia group and the time dependent model for the control and the
hyperthermia group. We used either adding/deleting or swapping moves
with equal probability at each step of the chain; we assigned a
probability of $\lambda= 0.5$ to the move that updates $\bof{R}$ and
then probability $0.25$ to each of the moves that update $\bof{R}'$
and $\bof{R}''$. In all cases, after the initial burn-in, both chains
mostly explored the same region of the parameter space corresponding
to configuration of $\bof{R}$ with high posterior probability. In
general, we found good agreement between the two chains, which were
run from different starting points. To be more precise, correlations
between the posterior probabilities of the two chains ranged from
$0.89$ to $0.91$.

\begin{figure}

\includegraphics{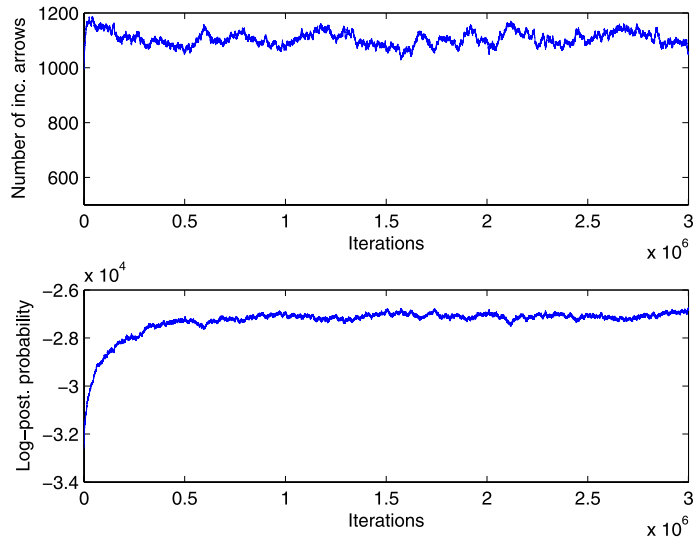}

\caption{Trace plot for number of selected arrows and for the
log-posterior probability for the time invariant model.}\label{trace}
\end{figure}

Figure \ref{trace} gives the summary trace plots for the number of
selected coefficients $\beta_{gm}$ and corresponding log-posterior
probabilities for the time invariant model on the hyperthermia group. In
this case the chain was run for three million iterations, from a
starting randomly chosen set of 1000 arrows, and mostly visited
models with 1000--1200 edges, that is, on average roughly 1 edge per
gene, a number not too far from the prior specification.

\subsection{Results}\label{sec:res}

The huge number of potential coefficients in the model implies that
the weight of a single coefficient toward the posterior probability
of the entire model can be potentially very small. Also, due to
sparsity, there may be many models with almost the same (small)
posterior probability. Because of this, it is good practice to
perform posterior inference based on the marginal posterior
probability of the single coefficients, rather than on their joint
distribution. These posterior probabilities of the presence of
single interactions, that is, $P(r_{gm}=1|\bof{Y},\bof{X})$, can be
estimated directly from the MCMC samples by taking the proportion of
MCMC iterations for which $r_{gm}=1$.

The small sample size of our experimental groups does not allow us
to create a validation set and, therefore, all the samples are used
to fit the model. Selected models are then evaluated based on the
$R^2$ statistic, calculated using the posterior mean of regression
coefficients. As expected, when more covariates are included into
the model, based on their posterior probabilities, the statistic
$R^2$ increases. We observed that coefficients with highest posterior
probability explain most of the variability,
while the increment in $R^2$ becomes marginal when adding coefficients with
relatively low posterior probability. We take this as an indication of
the fact that the ordering created by the
posterior probabilities correctly maps the significant
variables. For the time invariant model a threshold of 0.15,
corresponding to 1224 included edges, gave an $R^2$ of $0.36$, for
the control group, and of $0.44$ for the hyperthermia group, with
1473 included edges. Identical behavior was observed for the
additional coefficients of the time dependent model, that is,
when the number of included $\beta'$'s and $\beta''$'s increases,
then the quality of the fitting improves; with a threshold of $0.15$
for $\beta$'s and a threshold of $0.1$ for $\beta'$'s and
$\beta''$'s, we obtain a $R^2 = 0.45$ for the control group,
including 1826 $\beta$'s, 366 $\beta'$'s and 222 $\beta''$'s, and a
$R^2 = 0.47$ for the hyperthermia group, including 2173 $\beta$'s,
439 $\beta'$'s and 296 $\beta''$'s.

In an effort to assess whether our model correctly selects miRNAs
that under-regulate target genes, we also calculated the ordinary
least square estimates of the regression coefficients and checked
how many of them were negative; see the \hyperref[appendix]{Appendix} for the calculation of
the OLS estimates. Notice that this approach does not impose the
negative constraint on $\beta$'s. By including all coefficients with
posterior probability greater than 0.2 (0.15), we found that, for the
control and hyperthermia group, respectively, $96.4\,(93.5)\%$ and
$95.0\,(90.4)\%$ of the estimated coefficients were correctly
negative.

Important pairs of target genes and miRNAs can be selected as those
corresponding to arrows with highest posterior probabilities.
For example, by exploring the regulatory network as a function of this
posterior probability of the arrows, we found that, for the time
invariant model on the control group, a posterior probability
cutoff of 0.8 selected 88 arrows between 88 target genes and 7
miRNAs. These correspond to an expected rate of false detection
(Bayesian FDR) of 7.5\%, that we calculated, following
\citet{newton2004}, as
\[
\mathit{FDR} = \mathrm{C}(\kappa) / |J_{\kappa}|,
\]
where $\mathrm{C}(\kappa) = \sum_{g,m} \psi_{gm} I_{[\psi_{gm}\leq
\kappa]}$ and $\psi_{gm} = 1 - P(r_{gm}=1|\bof{Y},\bof{X})$, with
$|J_{\kappa}|$ the size of the list ($|J_{\kappa}| =\sum_{g,m}
I_{[\psi_{gm}\leq\kappa]}$). We set $\kappa=1-k$ with $k$ the
chosen threshold (i.e., 0.8). For the same cutoff the time-dependent
analysis on the control group showed 11 miRNAs with at least one target
gene for a total of 107 gene targets. There were 7 miRNA and 75 gene
targets in common between the time-independent and time-dependent
analyses of the control data.
For the hyperthermia-treated group, the time-invariant model with
a 0.8 cutoff led to 93 selected arrows, between 91 target genes
and 11 miRNAs, corresponding to a Bayesian FDR of 9.0\%.
The time-dependent analysis showed 12 miRNAs with at least one target
gene for a total of 120 gene targets.
There were 10 miRNA and 77 gene targets in common between the
time-independent and time-dependent analysis of the
hyperthermia-treated data.

\begin{figure}

\includegraphics{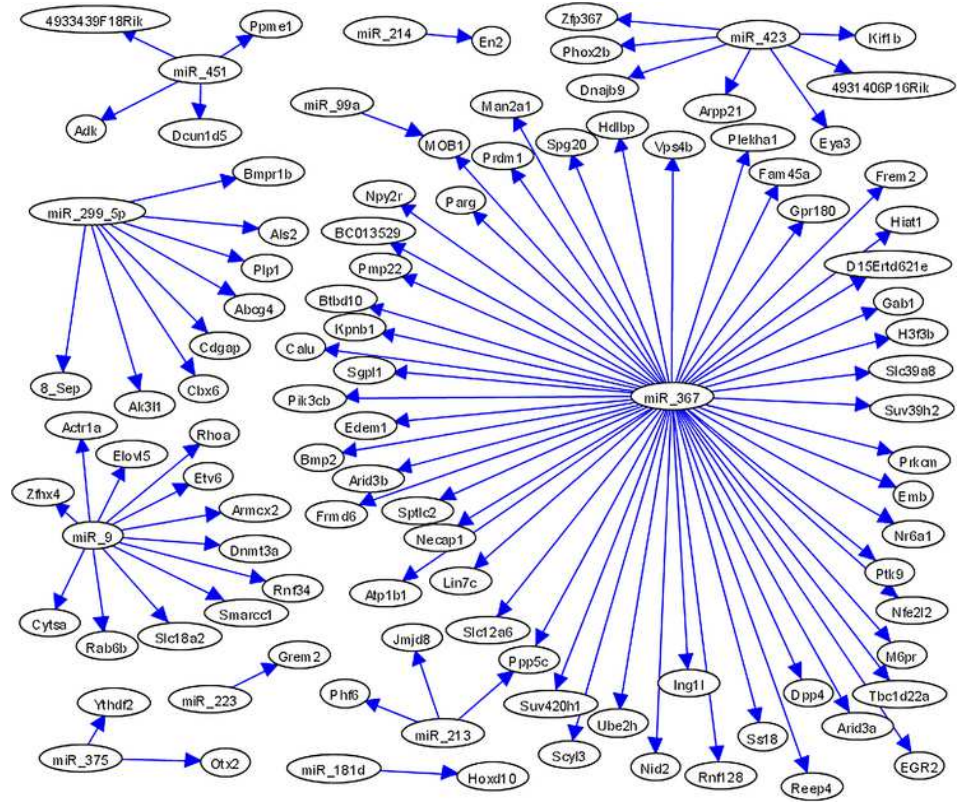}

\caption{Selected
network for the hyperthermia group using a threshold of 0.8 on the
posterior probability.} \label{NetPosterior}
\end{figure}

Figure \ref{NetPosterior}, produced using GraphExplore of
\citet{wang2004}, displays the selected network for the hyperthermia
group and a threshold of 0.8 on the posterior probability under
the time invariant model. A close look at the pairs of target genes
and miRNAs with high posterior probabilities reveals that some of
the regulatory relationships seem plausible and warrant future
investigation. For example, links between miR-367 and target Egr2 and
Mob1, selected with posterior probability of 0.97 and 1,
are particularly interesting, as described below.
%

Increasing the cutoff value on the posterior probabilities clearly
reduces the number of selected arrows and results in lower Bayesian
FDRs. For example, a cutoff value of 0.9 identified 60 arrows between
60 target genes and 6 miRNAs, corresponding to a Bayesian FDR of 4.1\%,
for the control group, and 50 selected arrows, between 50 target genes
and 9 miRNAs, corresponding to a Bayesian FDR of 3.9\%, for the
hyperthermia-treated group.

Overall, there were 252 unique miRNA-gene target associations
identified with a posterior probability of at least 0.8, including 15
miRNAs and 221 gene targets. Of the 252 miRNA-gene target associations
identified, 35 were predicted by miRanda only, 26 by PicTar only, zero
by total Target Scan only, 14 by aggregate Target Scan only, 8 by PITA
only, and 45 by at least one of the five algorithms considered. 108 of
the gene targets identified were associated with miR-367, a
pluripotency-specific marker in human and mouse ES cells [\citet
{li2009}], while 27 of the gene targets were associated with miR-423,
which has previously been shown to be expressed in the adult and/or
developing brain [\citet{Bzhang2009}]. Expression of both miR-367
and 423 decreased over time in control and hyperthermia-treated
embryos, which is consistent for a marker of pluripotency in a
differentiating embryo. While decreasing, the expression levels of
these miRNAs were higher after hyperthermia exposure when compared to
controls, which may indicate a delay in the differentiation program. In
addition, 62 of the gene targets were associated with miR-299-5p, which
has been shown to regulate de novo expression of osteopontin, a protein
that plays a role in enhancing proliferation and tumorigenicity
[\citet{shevde2009}].

Among the target genes identified, 12 genes associated with brain
development or expressed in brain/whole embryo (including Egr2, Hnf1b,
and Mob1) were associated with miR-367 with a posterior probability in
0.82--1.0 in the time-dependent analysis of hyperthermia-treatment data
at the 5-hour time point. 11 of these gene target associations also had
a posterior probability in 0.68--1.0 with the time-independent analysis
of hyperthermia treatment data. At the 5-hour time point after
hyperthermia treatment, miR-367 expression increased 1.7-fold, while
expression of these associated gene targets decreased 1.1--2 fold when
compared to control-treatment. This pattern of expression might be
indicative of down-regulation of these gene targets by increased
expression of miR-367 in response to hyperthermia treatment. Such
gene-miRNA associations, identified by our methods as possibly related
to brain and embryo development, are interesting to pursue in follow-up
NTD studies.

\begin{figure}

\includegraphics{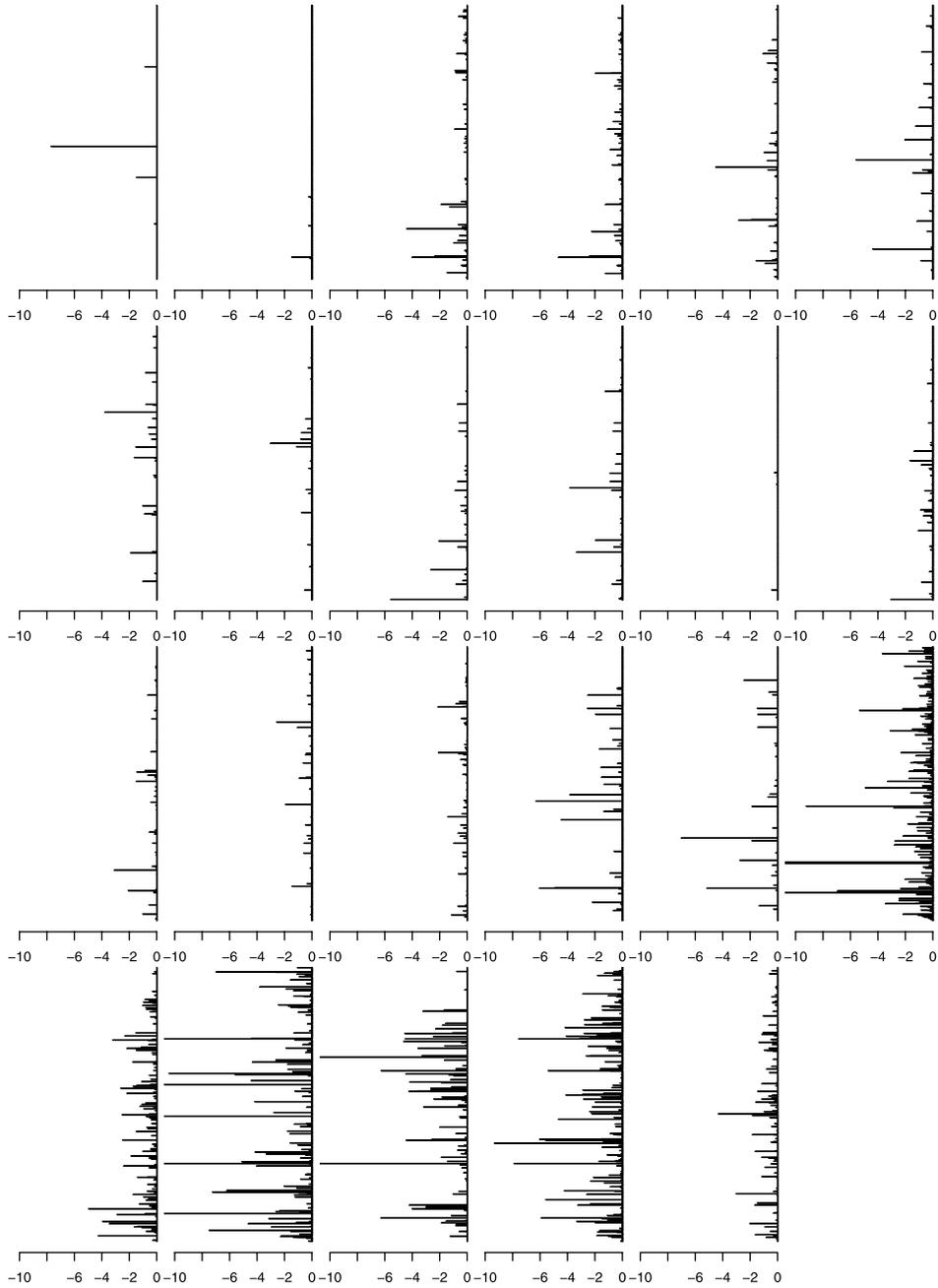}

\caption{Estimation of the significant $\beta_{gm}$'s for the time
invariant model under hyperthermia condition.
The y-axis indicates the 1297 targets, listed in the same order in all
23 plots.}\label{barBeta}
\end{figure}

It is also interesting to look at the inference on the regression
coefficients. Figure \ref{barBeta} shows the estimates of the
significant $\beta_{gm}$'s for the time invariant model under
hyperthermia condition. Each bar in the plot represents the 1297
regression coefficients for one of the 23 miRNAs. Nonzero values
correspond to the posterior mean estimates of the best
$\beta_{gm}$'s with posterior inclusion probability above 0.15 (all
other $\beta$'s are estimated by zero). Notice, for example, how
miRNAs miR-423, corresponding to the $20${th} bar, and miR-367,
corresponding to the $18${th} bar, play an important role in the
down-regulatory mechanism.

\begin{figure}

\includegraphics{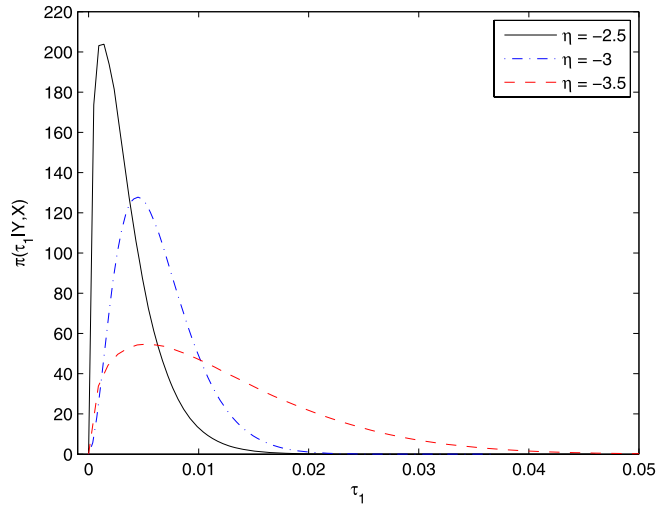}

\caption{Density Kernel estimate using the time independent model
for the control group.} \label{figTau}
\end{figure}

Let us now comment on the inference on $\tau_1, \ldots, \tau_5$.
These parameters measure the influence of the prior information on the
posterior inference. In general, we noticed that posterior inference on
these parameters showed some sensitivity to the value assigned to $\eta
$. When selecting edges the hyperparameter $\eta$ represents the
weight assigned to the data and, consequently, $\tau_1, \ldots, \tau
_5$ play the role of the weight of the prior sequence information
derived from the five used algorithms. The bigger the value of $\eta$,
the more the posterior distribution of $\tau_j$ will be concentrated
around small values. Besides this general rule, inference on the $\tau
_j$'s generally depends on the concordance between data and prior
information, the number of observations and the number of parameters in
the model. As an example, the behavior of the posterior distribution of
$\tau_1$ (the parameter associated to PicTar), for different values of
$\eta$, is summarized in Figure \ref{figTau}. The scale of the
estimates compensates the very large values we observe for some of the
PicTar scores. We can clearly see how the posterior distribution
concentrates on bigger values when $\eta$ decreases. To evaluate the
influence of the different scores, we calculated how the prior
probabilities (\ref{bernprior}) increase, on average, for a set of
pairs target-miRNA with high scores. With $\eta=-3.5$ the prior
probability of $r_{gm}=1$ increase by 35.9\%, while when setting $\eta
=-2.5$ this increment is equal only to 8.7\%.

\begin{figure}

\includegraphics{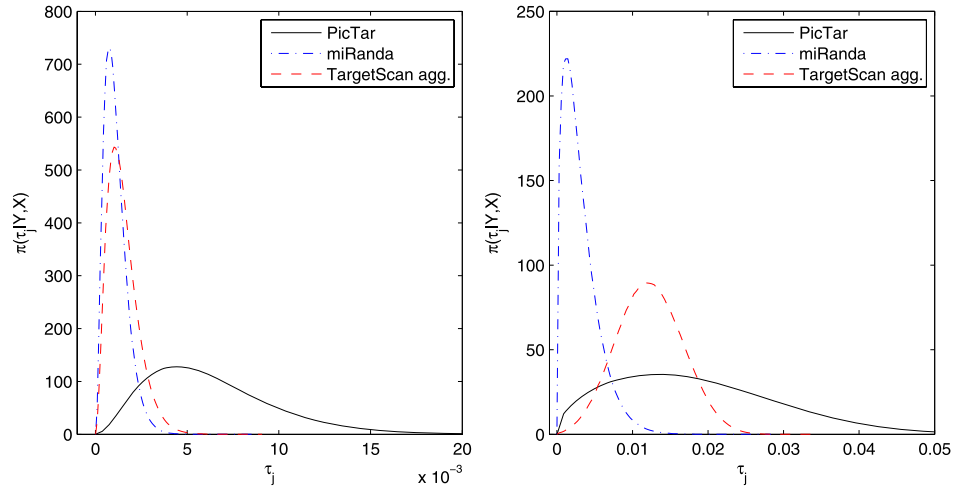}

\caption{Density Kernel estimate using the time independent model for
the control group (left panel) and the time dependent model
for the hyperthermia group (right panel).} \label{figTauConfronto}
\end{figure}

Our results suggested that information extracted from PicTar is the
most influential on the posterior inference. MiRanda and Target Scan
aggregate also contribute somehow to the selection process, while
Target Scan total and PITA do not affect the posterior inference.
Figure \ref{figTauConfronto} shows the posterior inference for the
three most influential algorithms when $\eta=-3$. The other two
algorithms resulted in posterior densities that were very concentrated
around zero (plot not shown). Notice that, while the importance order
does not change, the 3 algorithms are generally more influential in the
case of the time dependent model for the hyperthermia group (right
panel). Also, Target Scan aggregate seems to have an increased effect
for this case, compared to the model with time invariant coefficients
on the control group (left panel). In general, we found that with $\eta
=-3$ the posterior distributions of $\tau_1, \ldots, \tau_5$, for
the control group, are concentrated around values that imply a 16.4\%
increase on the prior probability of $r_{gm}=1$, for high scores. For
the hyperthermia group the corresponding percentage, as consequence of
the increased influence of PicTar and Target Scan aggregate, increases
to 37.8\%. When using the time dependent model the prior probability of
$r_{gm}=1$ increases by 32.6\% for the control group and, due mostly to
the increased effect of PicTar and Target Scan aggregate, by 118.2\%
for the hyperthermia group.

Because different sequence/structure based methods do not result in
exactly the same set of miRNA target predictions, we decided to
perform a systematic analysis to understand how different prior
target predictions can influence the final inference. Accordingly,
we selected the control group and the model with time invariant
coefficients and repeated our analysis by integrating in the prior
formulation PicTar, miRanda, Target Scan aggregate, Target Scan
total and PITA scores, one set at the time. We then compared the
selected arrows obtained integrating single sets to those selected
by integrating all five available sets of scores. Using $\eta= -3$,
we found that the network selection is only partially affected by
the different data integrations. In particular, by considering the
first 200 arrows, selected according to the posterior probability,
we found that the set of selected ones using PicTar overlaps at the
73.5\% with the set selected using all 5 databases; this percentage
is equal to 66.0\% for miRanda, to 75.0\% for Target Scan aggregate,
to 70.0\% for Target Scan total and 72.5\% for PITA. If we instead
consider the first 1000 arrows, then these percentages get lower
(66.4\% for PicTar, 60.0\% for miRanda, 68.8\% for Target Scan
aggregate, 65.8\% for Target Scan total and 68.2\% for PITA). These
results indicate that although the selection is mostly data driven,
especially when restricted to the most important arrows, it is also
partially affected by the integrated sets.

\section{Conclusions}\label{sec:final}

We have proposed a Bayesian graphical modeling approach that infers
the miRNA regulatory network by integrating expression levels of
miRNAs with their potential mRNA targets and, via the prior
probability model, with their sequence/structure information. Our
model is able to incorporate multiple data sources directly into the
prior distribution, avoiding arbitrary prior data synthesis. We have
used stochastic search variable selection methods to infer the miRNA
regulatory network. We have considered
experimental data from a study on a very well-known developmental
toxicant causing neural tube defects, hyperthermia. The analysis has
involved 23 mouse miRNAs and a total of 1297 potential targets. Our
goal was to identify a small set of potential targets with high
confidence. Some of the pairs of target gene and miRNA selected by
our model seem promising candidates for future investigation.
In addition, the time-dependent model has achieved
significant improvement in the percentage of explained variance, only
slightly increasing the size of the selected model. Our
proposed modeling strategy is general and can easily be applied to
other types
of network inference by integrating multiple data sources.

An interesting feature of our inference is that there is only a partial
overlap between the connections selected by our model and those
indicated by the sequence/structure algorithms.
This phenomenon has been observed by other authors in models for data
integration.
\citet{wei2008}, for example, attribute this to the fact that
our knowledge of biological processes is not complete and can potentially
include errors and therefore induce misspecified edges on the networks.
They also suggest to first check the consistency of the prior information
with the available data. In our case, if the correlation between a miRNA
and a target gene is very small, we may want to remove the
edge from the network. On the other hand, given the limited number of
observations
typical of experimental studies in genomics, it would seem important to retain
as much, possibly accurate, prior information as possible.
This important aspect of models for data integration certainly deserves
future investigation.

Extensions and generalizations of our model are possible. One future
avenue we intent to pursue is trying to relax the assumption on the
conditional independence of the targets given the miRNAs. This
assumption is necessary in order to integrate out the covariance
matrix, as in \citet{brown1998}, and still allow the selection of
individual relations between a gene and a miRNA. Looking at this as a
computational issue, it may be possible to still sample the values of
this huge covariance matrix in the MCMC, perhaps by reducing the number
of nonzero elements via the prior information on the gene network.

\begin{appendix}
\section{Posterior inference on regression coefficients}\label{appendix}
If inference on regression coefficients is desirable, these can
estimate either via the posterior distributions or the least squares
estimates. For model (\ref{model}) we have the following posterior
distribution:
\begin{equation}
\pi\bigl(\beta_g |\mathbf{Y}, \mathbf{X}_{(R)}, \omega^2\bigr)  \sim  HN^{+}(U_g C_g, \sigma_g U_g), \\
\end{equation}
where $HN^{+}$ indicates a $k_g$-variate half-normal distribution
that gives positive probability only to vectors formed by elements
bigger than zero.

For the more general time-dependent model we have the following
posterior distributions:
\begin{equation}
\cases{
\pi\bigl(\beta_g |\mathbf{Y}, \mathbf{X}_{(R)}, \omega^2\bigr)  \sim  HN^{+}(E_g^{-1} F_g, \sigma_g E_g^{-1}),\cr
\pi\bigl(\beta_g'' |\mathbf{Y}, \mathbf{X}_{(R)}, \omega^2\bigr)  \sim N(J_g^{-1}H_g, \sigma_g J_g^{-1}),
}
\end{equation}
with
\begin{eqnarray*}
J_g &=& \mathbf{X}^{T}_{3(R'')} \mathbf{X}_{3(R'')} - \mathbf{X}_{3
(R'')}^{T} \mathbf{X}_{3(R)} L_g^{-1} \mathbf{X}_{3 (R)}^{T}
\mathbf{X}_{3(R'')} + (\zeta c)^{-1} I_{k_{3g}}, \\
H_g & = &  \mathbf{Y}_{3g}^T \mathbf{X}_{3(R'')}\\
&&{} +\bigl(\mathbf{Y}_g^T
\mathbf{X}_{(R)} - \mathbf{Y}_{2g}^T \mathbf{X}_{2(R')} D_g^{-1}
\mathbf{X}_{2(R')}^T \mathbf{X}_{2(R)} + \sigma_g^{1/2}c^{-1}
\mathbf{1}_{k_g}\bigr)L_g^{-1}\mathbf{X}_{3 (R)}^{T} \mathbf{X}_{3 (R'')}, \\
D_g & = & \mathbf{X}_{2(R')}^{T} \mathbf{X}_{2(R')} + (\zeta c)^{-1} I_{k_{2g}}, \\
L_g & = & \mathbf{X}_{(R)}^{T} \mathbf{X}_{(R)}
-\mathbf{X}_{2(R)}^{T} \mathbf{X}_{2(R')} D_g^{-1}
\mathbf{X}_{2(R')}^{T} \mathbf{X}_{2(R)}. \\
\end{eqnarray*}
The posterior distribution of $\beta'$ has the same form
as the posterior distribution of~$\beta''$. Using the least squares
approach, instead, we obtain the following equations for $\beta$,
$\beta'$ and $\beta''$:
\begin{eqnarray*}
\cases{
\hat{\beta}_{gLS} = \bigl(\mathbf{X}_{(R)}^{T} \mathbf{X}_{(R)}\bigr)^{-1}\mathbf{X}_{(R)}^{T} \bigl(Y_g - \mathbf{X}_{2(R')}^* \beta_{g}' +\mathbf{X}_{3(R'')}^* \beta_{g}''\bigr), \cr
\hat{\beta'}_{gLS}  =  \bigl(\mathbf{X}_{2(R')}^{T}\mathbf{X}_{2(R')}\bigr)^{-1}\mathbf{X}_{2(R')}^{T} \bigl(Y_{2g} - \mathbf{X}_{2(R)} \beta_{g}\bigr), \cr
\hat{\beta''}_{gLS}  =  \bigl(\mathbf{X}_{3(R'')}^{T}\mathbf{X}_{3(R'')}\bigr)^{-1}\mathbf{X}_{3(R'')}^{T} \bigl(Y_{3g} - \mathbf{X}_{3(R)} \beta_{g}\bigr),
}
\end{eqnarray*}
and then
\begin{eqnarray*}
&&\hat{\beta}_{gLS}  =  K_g^{-1} \bigl[ \hat{\beta}_{gOLS}-\bigl(\mathbf{X}_{(R)}^{T} \mathbf{X}_{(R)}\bigr)^{-1} \mathbf{X}_{(R)}^{T}
\bigl(\mathbf{X}_{2(R')}^* \bigl(\mathbf{X}_{2(R')}^{T}
\mathbf{X}_{2(R')}\bigr)^{-1}\mathbf{X}_{2(R')}^T
Y_{2g} \\
&&\hspace{164pt} {}+ \mathbf{X}_{3(R'')}^* \bigl(\mathbf{X}_{3(R'')}^{T}
\mathbf{X}_{3(R'')}\bigr)^{-1}\mathbf{X}_{3(R'')}^T
Y_{3g}\bigr) \bigr], \\
\end{eqnarray*}
 with
\begin{eqnarray*}
&&K_g  =   I_{k_g} - \bigl(\mathbf{X}_{(R)}^{T} \mathbf{X}_{(R)}\bigr)^{-1}
\mathbf{X}_{(R)}^{T} \bigl(\mathbf{X}_{2(R')}^* \bigl(\mathbf{X}_{2(R')}^{T}
\mathbf{X}_{2(R')}\bigr)^{-1}\mathbf{X}_{2(R')}^T
\mathbf{X}_{2(R)} \\
&&\hspace{136pt}{} + \mathbf{X}_{3(R'')}^* \bigl(\mathbf{X}_{3(R'')}^{T}
\mathbf{X}_{3(R'')}\bigr)^{-1}\mathbf{X}_{3(R'')}^T
\mathbf{X}_{3(R)} \bigr).
\end{eqnarray*}

\end{appendix}
\section*{Acknowledgments}
The authors would like to
acknowledge discussions with David Dahl and Adarsh Joshi on a
preliminary modeling approach to the data used in this paper.


\printaddresses

\end{document}